# Temperature-aware Dynamic Optimization of Embedded Systems

TOSIRON ADEGBIJA, University of Arizona
ANN GORDON-ROSS, University of Florida

Due to embedded systems' stringent design constraints, much prior work focused on optimizing energy consumption and/or performance. Since embedded systems typically have fewer cooling options, rising temperature, and thus temperature optimization, is an emergent concern. Most embedded systems only dissipate heat by passive convection, due to the absence of dedicated thermal management hardware mechanisms. The embedded system's temperature not only affects the system's reliability, but could also affect the performance, power, and cost. Thus, embedded systems require efficient thermal management techniques. However, thermal management can conflict with other optimization objectives, such as execution time and energy consumption. In this paper, we focus on managing the temperature using a synergy of cache optimization and dynamic frequency scaling, while also optimizing the execution time and energy consumption. This paper provides new insights on the impact of cache parameters on efficient temperature-aware cache tuning heuristics. In addition, we present temperature-aware phase-based tuning, TaPT, which determines Pareto optimal clock frequency and cache configurations for fine-grained execution time, energy, and temperature tradeoffs. TaPT enables autonomous system optimization and also allows designers to specify temperature constraints and optimization priorities. Experiments show that TaPT can effectively reduce execution time, energy, and temperature, while imposing minimal hardware overhead.

## 1. INTRODUCTION

Embedded systems have become ubiquitous over the past few years, and with the emergence and growth of the Internet of Things, embedded systems are expected to become even more pervasive. Researchers have focused on effective optimization techniques for optimizing embedded systems' energy consumption, since these systems typically have stringent resource and design constraints [Gordon-Ross and Vahid 2003; Hajimir and Mishra 2012]. These constraints include form factor, battery capacity, cost, real-time deadlines, etc., and pose significant challenges to embedded system optimization. The optimization challenges are exacerbated by the increase in high-demand (compute/memory intensive) applications that must be executed within these resource constraints. Since most embedded systems are battery operated, much research efforts have focused on reducing energy consumption without significantly degrading system performance. However, temperature is also a growing issue in embedded systems optimization research since most embedded systems have fewer cooling options as compared to general purpose computers due to area/size, cost, and energy constraints. Most embedded systems lack traditional cooling mechanisms, such as active cooling fan, water cooling, heat pipes/sinks, etc., and only dissipate heat by passive convection. These

Author's addresses: T. Adegbija, Department of Electrical and Computer Engineering, University of Arizona, PO Box 210104, Tucson AZ, 85721, USA. Email: tosiron@email.arizona.edu

A. Gordon-Ross, Department of Electrical and Computer Engineering, University of Florida, PO Box 116200, Gainesville FL, 32611, USA. Email: ann@ece.ufl.edu

A. Gordon-Ross is also with the NSF Center for High-Performance Reconfigurable Computing.

constraints necessitate efficient thermal management techniques that impose minimal hardware overhead.

An embedded system's temperature affects several optimization goals, such as performance, reliability, power, and system cost. Increased chip temperature in an embedded system can increase cooling costs, and reduce performance, mean time to failure (MTTF), and reliability. In addition, increased temperature can lead to thermal emergencies, which can result in an exponential increase in leakage power and thermal runaway, leading to permanent chip damage. To address these issues, several dynamic thermal management (DTM) techniques have been proposed for managing chip temperature. Most of these techniques leverage clock gating [Brooks and Martonosi 2001], dynamic voltage scaling (DVS), dynamic frequency scaling (DFS), or dynamic voltage and frequency scaling (DVFS) [Skadron 2004], and/or task migration [Heo et al. 2003].

In our work, we use DFS as part of a broader technique for thermal management in embedded systems. DFS is an effective dynamic thermal optimization technique that adjusts a microprocessor's frequency to changing application resource requirements, thereby reducing the microprocessor's power consumption and/or heat dissipation. DFS is commonly implemented in modern day microprocessors [ARM 2016], especially in battery-operated/resource-constrained devices, such as smartphones. The frequency at which the circuit is clocked determines the voltage required for stable operation, therefore, the voltage can be reduced as the frequency is reduced. Thus, DFS is commonly used in conjunction with DVS, and is sometimes referred to as DVFS. While we explicitly utilize DFS in our work, the work presented herein is also applicable to DVS or DVFS.

One of the potential drawbacks of DTM techniques is that optimizing the temperature in isolation can significantly degrade other optimization goals, such as execution time and/or energy consumption [Pedram and Narian 2006]. In addition, the applications' execution characteristics (e.g., cache misses, instruction per cycle (IPC), branch mispredictions, etc.) can also affect the temperature [Inchoon and Kim 2009]. Since applications typically have dynamically varying execution characteristics, we show that the temperature can be further optimized by considering intra-application characteristic variations. Some previous DTM techniques (e.g., [Jayaseelan and Mitra 2008]) consider inter-application characteristic variations, however, in this work, we optimize the system at a finer granularity than most previous works by considering intra-application characteristic variations.

To increase optimization potential by specializing system resources to varying application characteristics, we leverage phase-based tuning [Gordon-Ross et al. 2008] as a complementary approach to DFS. A phase is a length of execution where an application's characteristics remain relatively stable, and therefore the best system configuration, or specific parameter values (e.g., cache size, associativity, line size, clock frequency, etc.), that adhere to the design constraints also remain relatively stable. Phase-based tuning requires configurable hardware with tunable parameters, whose values can be specified/changed during runtime. Phase-based tuning also requires a mechanism to evaluate the application's characteristics in order to determine the best system configurations that satisfy each phase's resource requirements. Previous work showed that phase-based tuning significantly reduced energy consumption in embedded systems [Gordon-Ross et al. 2008]. For example, Gordon-Ross et al. [Gordon-Ross et al. 2005] showed that phase-based cache tuning reduced the cache's memory access energy by up to 62%. However, few studies have addressed the combination of phase based-tuning and DTM.

In order to maximize the benefits of phase-based tuning and DTM, the tuned system components must be carefully selected. In this work, we focus on the cache for phase-based tuning, and the clock frequency for DTM, using DFS. On-chip caches are well known to account for a significant portion of a microprocessor's total energy consumption [Khaitan and McCalley 2014, Zhang et al. 2003]. In addition, caches could also be a performance bottleneck, since they are used to bridge the processor-memory performance gap. However, even though previous work has shown that caches contribute significantly to a chip's temperature [Huang et al. 2004], the thermal impacts of cache configurations have not been thoroughly investigated.

In this paper, we thoroughly investigate and analyze the thermal impacts of cache configurations and use the insight from this analysis to develop a low-overhead and flexible optimization heuristic that optimizes the temperature without degrading the execution time and/or energy. For the first time, to the best of our knowledge, we establish the impact order of cache parameters on the system temperature. This impact order will drive future advances in efficient temperature-aware cache tuning heuristics/algorithms.

We present a dynamic optimization heuristic—temperature-aware phase-based tuning (TaPT), which dynamically determines the Pareto optimal system configurations trading off execution time, energy, and temperature design objectives. TaPT is based on the strength Pareto evolutionary algorithm II (SPEA2) [Zitzler et al. 2001], which is a well-known and effective evolutionary algorithm for solving multi-objective optimization problems. We modify SPEA2 to implement phase-based tuning and consider designer-selected priority settings. These priority settings offer designers the flexibility to choose which design objective to prioritize during optimization. TaPT's runtime automation aids designers in adhering to design constraints with minimal design time effort. TaPT leverages previously proposed/existing configurable hardware, thus minimizing the additional hardware overhead with respect to these prior techniques. Experimental results show that compared to using the same system configuration throughout an application's execution, TaPT reduces execution time, energy consumption, and temperature by up to 5%, 30%, and 25%, while adhering to designer-specified design constraints. Additionally, we compare TaPT to DFS and cache tuning in isolation, and quantitatively illustrate the benefits and tradeoffs of TaPT over DFS and cache tuning. Finally, we show that TaPT can be easily implemented, requires minimal design time effort, and constitutes minimal hardware overhead with respect to state-of-the-art embedded system microprocessors.

## 2. BACKGROUND AND RELATED WORK

Much previous work focused on phase-based tuning [Adegbija and Gordon-Ross 2014; Gordon-Ross et al. 2008; Hajimir and Mishra 2012] and DTM [Brooks and Martonosi 2001; Heo et al. 2003; Salami 2014; Skadron 2004]. Since we leverage both phase-based tuning and DTM, we present related work and background in these two areas. We also present background and key concepts from SPEA2 that we leveraged for TaPT.

### 2.1 Phase-based Tuning and DTM

To facilitate phase-based tuning, hardware- or software-based phase classification partitions an application's execution into intervals, measured by the number of instructions executed. Intervals showing similar characteristics can be clustered into phases. Sherwood et al. [Sherwood and Calder 1999] studied applications' time varying behaviors using SPEC 95 benchmarks, and showed that applications have periodic patterns and exhibit phase-based behavior with respect to several execution statistics (e.g., cache miss rates, branch mispredicts, IPC, etc.) Balasubramonian et al. [Balasubramonian 2000] used cache miss rates, cycles per instruction (CPI), and branch frequency characteristics to detect changes in application characteristics for cache tuning, and found that these characteristics were effective for phase classification. Dhodapkar et al. [Dhodapkar and Smith 2001] found a relationship between phases and the phases' working sets, and concluded that phase changes could be detected by detecting changes in the working set. In this work, we use execution statistics obtained from the microprocessor's hardware performance counters for phase classification [Sembrant et al. 2011]. Since we utilize cache tuning in this work, for brevity, we limit our review to phase-based cache tuning.

A major challenge in phase-based tuning is tuning the configurable hardware to the best configuration for each phase without incurring significant tuning overhead. Zhang et al. [Zhang et al. 2003] proposed a cache tuning heuristic that traded off energy consumption and performance to determine the Pareto optimal cache configurations. The heuristic searched the cache parameters in order of the parameters' impact on the energy consumption, first determining the best cache size, followed by the best line size, and finally the best associativity. However, this method executed several inferior, non-optimal configurations, thus incurring tuning overhead. Gordon-Ross et al. [Gordon-Ross et al. 2008] presented cache design space exploration heuristics that when used for phased-based tuning, realized as much as 39% energy savings on average as compared to non-phase-based tuning (i.e., using a single configuration for the entire application). Hajimir et al [Hajimir and Mishra 2012] presented a dynamic programming-based algorithm to find the best cache configuration for each phase. However, these methods only focused on energy savings and did not consider thermal issues.

Huang et al. [Huang et al. 2004] showed that the cache contributes significantly to the overall chip temperature, and necessitates optimization techniques that target cache thermal management. Homayoun et al. [Homayoun et al. 2012] showed that the memory cell peripherals' power dissipation is significantly higher than the actual memory cell's power dissipation. This difference in power dissipation results in thermal variations within the cache, and is caused by the difference in activity factors between peripheral logic and memory cells. Peripheral logic includes global and local address routing drivers, global data in/out drivers, row predecoder drivers, and wordline drivers. Additionally, the different types of transistors used for the peripheral logic and memory cell also contribute to the cache's thermal variation.

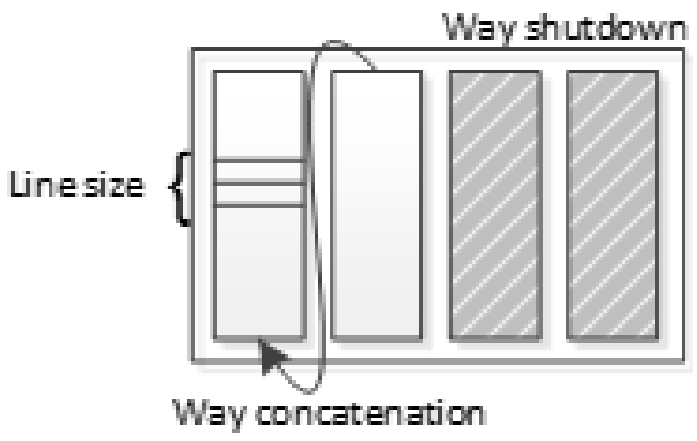

Fig. 1. Configurable cache architecture.

To reduce chip temperature dissipation, several DTM techniques have been proposed. Brooks et al. [Brooks and Martonosi 2001] investigated clock gating, which turns off the clock signals during thermal emergencies. Heo et al. [Heo et al. 2003] proposed task migration, which migrated tasks from a hot core to a cooler core to avoid a thermal emergency. More recently, Liu et al. [Liu et al. 2013] proposed a DTM technique that used task migration to reduce temperature variations across the chip, while considering transient thermal effects. Kong et al. [Kong et al. 2012] presented a survey of recent thermal management techniques for microprocessors, focusing on the techniques that affect or rely on the microarchitecture. The authors showed that most DTM techniques have the potential to degrade performance due to longer execution times. Additionally, these works did not explicitly consider the tradeoffs between energy, temperature, and execution time, thus increasing the possibility of significantly degrading one design objective while optimizing other design objectives. Furthermore, these methods were not phase-based and did not consider intra-application/intra-task variations.

Since prior work showed that phase-based cache tuning significantly impacts energy consumption and execution time, and DTM techniques can significantly impact temperature, energy consumption, and execution time, we combine phase-based cache tuning and DFS for fine-grained and efficient temperature, energy, and execution time optimization. However, since optimizing one design objective may adversely impact the other design objectives, combining these techniques presents a multi-objective optimization problem. The solution to a multi-objective optimization problem is the Pareto optimal configuration set, which enables designers to choose the system configuration that best meets the design constraints.

Our work improves the robustness of thermal management and explores the synergies between different optimization techniques. We combine phase-based cache tuning and DFS to determine Pareto optimal configurations that trade off execution time, energy, and temperature, thus increasing optimization potential and achieving fine-grained multi-objective optimization.

### 2.2 Configurable Hardware

Phase-based tuning can leverage any configurable cache architecture (e.g., [Gordon-Ross and Vahid 2003]) and tuning method to search the configuration design space, which consists of all the different system configurations/combinations of tunable parameter values. Motorola's M*CORE processor [Motorola 2014] provided per-way configuration using way management, which allowed ways to be shut down or designated as instruction only, data only, or unified. For our work, we assume highly configurable, private, separate level one (L1) instruction and data caches, however our methods could be extended to consider additional levels of cache. Fig. 1 depicts our configurable cache architecture, which is based on the configurable cache proposed by Zhang et al. [Zhang et al. 2003]. This configurable cache provides runtime-configurable total cache size, associativity, and line size using a small, hardware-settable bit-width configuration register. Configurable associativity is achieved by logically concatenating ways, configurable size is achieved by shutting down ways, and configurable line size is achieved by fetching additional physical cache lines for larger, logical line sizes. We elaborate on the achievable design space given this configurability in Section 4.1.

### 2.3 SPEA2 Algorithm

Evolutionary algorithms leverage biological evolutionary concepts, such as population, reproduction, mutation, selection, etc., for efficiently determining Pareto optimal solutions to multi-objective optimization problems. The solution space consists of all of the possible solutions to the optimization problem, the population is a subset of the solution space, and the population's solutions are referred to as individuals. A solution's fitness dictates the solution's quality and represents how well the solution adheres to design constraints. Evolution iterates over successive generations of population, where each evolution considers the population's individuals' finesses and replaces the least fit individuals with new solutions from the solution space, and interjects random solution mutations, to create the successive generation.

Prior work shows that SPEA2 outperforms most other evolutionary algorithms for solving multi-objective optimization problems [Zhang et al. 2003]. SPEA2 uses *elitism,* which maintains an external set of non-dominated solutions, called an archive. A solution is non-dominated (or Pareto optimal) if none of the design objectives can be improved without degrading another design objective. For example, given two configurations $C_x$ and $C_y$, $C_x$ dominates $C_y$ (written as $C_x \succ C_y$) if and only if:

$$\forall i \in \{1, 2, \ldots, k\} : f_i(C_x) \geq f_i(C_y) \quad \exists j \in \{1, 2, \ldots, k\} : f_j(C_x) > f_j(C_y) \qquad (1)$$

where k is the number of objectives and $f_k$ represents the design objectives' objective functions, and $f_k(C_x)$ characterizes how well Cx achieves the design objectives.

For brevity, we present an overview of SPEA2, and refer the reader to [Zhang et al. 2003] for additional details. SPEA2 takes the solution space as input and outputs the Pareto optimal solution set. SPEA2 generates an initial population and creates an empty archive and populates the first generation's archive with the population's non-dominated individuals. For subsequent generations, SPEA2 calculates the population's and archive's individuals' finesses, and populates the next generation's archive with the population's and archive's non-dominated individuals. When the maximum number of generations has been reached and/or number of solutions that satisfy the design objectives have been determined, the current archive contains the Pareto optimal set.

### 3. TEMPERATURE-AWARE PHASE-BASED TUNING (TAPT)

Previous research [Kong et al. 2012] on power and thermal management techniques have showed an intersection between power and thermal management hardware mechanisms. Most hardware mechanisms for power management can be leveraged for thermal management, since power reduction can also lead to temperature reduction. Thus, thermal management need not be a complex or high-overhead process, since the microprocessor designers typically do not need to adopt additional hardware specifically for thermal management. Since hardware mechanisms that can be leveraged for power management are commonly available in modern-day microprocessors, these mechanisms can also be leveraged for thermal management with new algorithms and management policies. However, thermal/power management techniques must consider and limit incurred performance degradations and/or increases in energy consumption.

TaPT leverages several fundamental assumptions based on mechanisms that have been widely studied and implemented in embedded systems [Karan et al. 2009; Zhang et al. 2003]. Fig. 2 depicts TaPT's synergistic interactions with different system components to dynamically optimize the system. We assume that DFS is enabled, and the system features a configurable cache with tunable size, associativity, and line size. We also assume that the system has temperature sensors that can be read by a continuous system telemetry mechanism for collecting and analyzing sensor data. Most current microprocessors contain hardware performance counters that generate execution statistics, such as cache miss rates, instructions per cycle, etc. These statistics are used in combination with low-overhead analytical models to estimate the power/energy consumption and performance [Karan et al. 2009], which are used by the TaPT characterization algorithm for determining the best configurations.

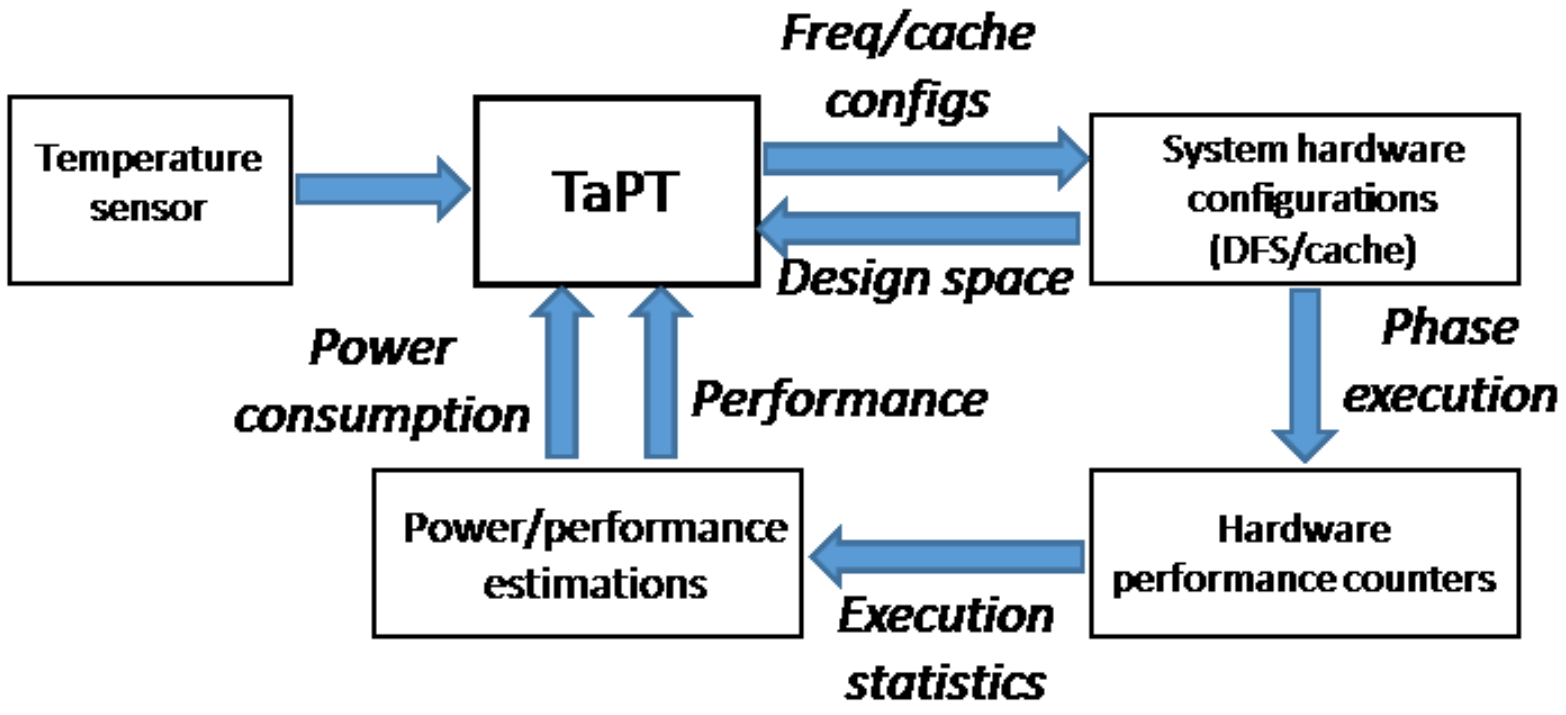

Fig. 2. TaPT interactions with other system components and functions

TaPT can be implemented as a software subroutine using the system's microprocessor, which enables easy system integration with state-of-the-art microprocessors. However, a software implementation can affect the system cache and applications' runtime behaviors due to context switching. These effects can cause TaPT to choose non-optimal, inferior configurations. Alternatively, TaPT can be implemented using non-intrusive, low-overhead custom hardware, with minimal negative impact on the system's area, energy, and performance. Due to the advantages of the hardware implementation, we assume the hardware approach for implementing TaPT. In this section, we present an overview of TaPT and details of the TaPT architecture and algorithm.

### 3.1 TaPT Overview

Fig. 3 depicts an overview of TaPT. When an application is executed, TaPT determines whether or not the application has been previously characterized (i.e., the best configurations for the application's phases have been determined). If the application is new, TaPT classifies the application's phases. TaPT classifies the phases by montoring application execution on the base configuration, during which the application execution statistics (e.g., IPC, instruction and data cache miss rates, etc.) are gathered at tuning intervals from the microprocessor's hardware performance counters. The tuning interval can be measured in number of executed instructions (e.g., 1 million instructions) or in time. For our experiments, we used tuning intervals of 10 ms, which we empirically determined to be sufficient to gather stable execution statistics. Execution lengths with similar execution characteristics are clustered to form phases, and these phases' characteristics are then used in the rest of the algorithm to determine the phases' best configurations.

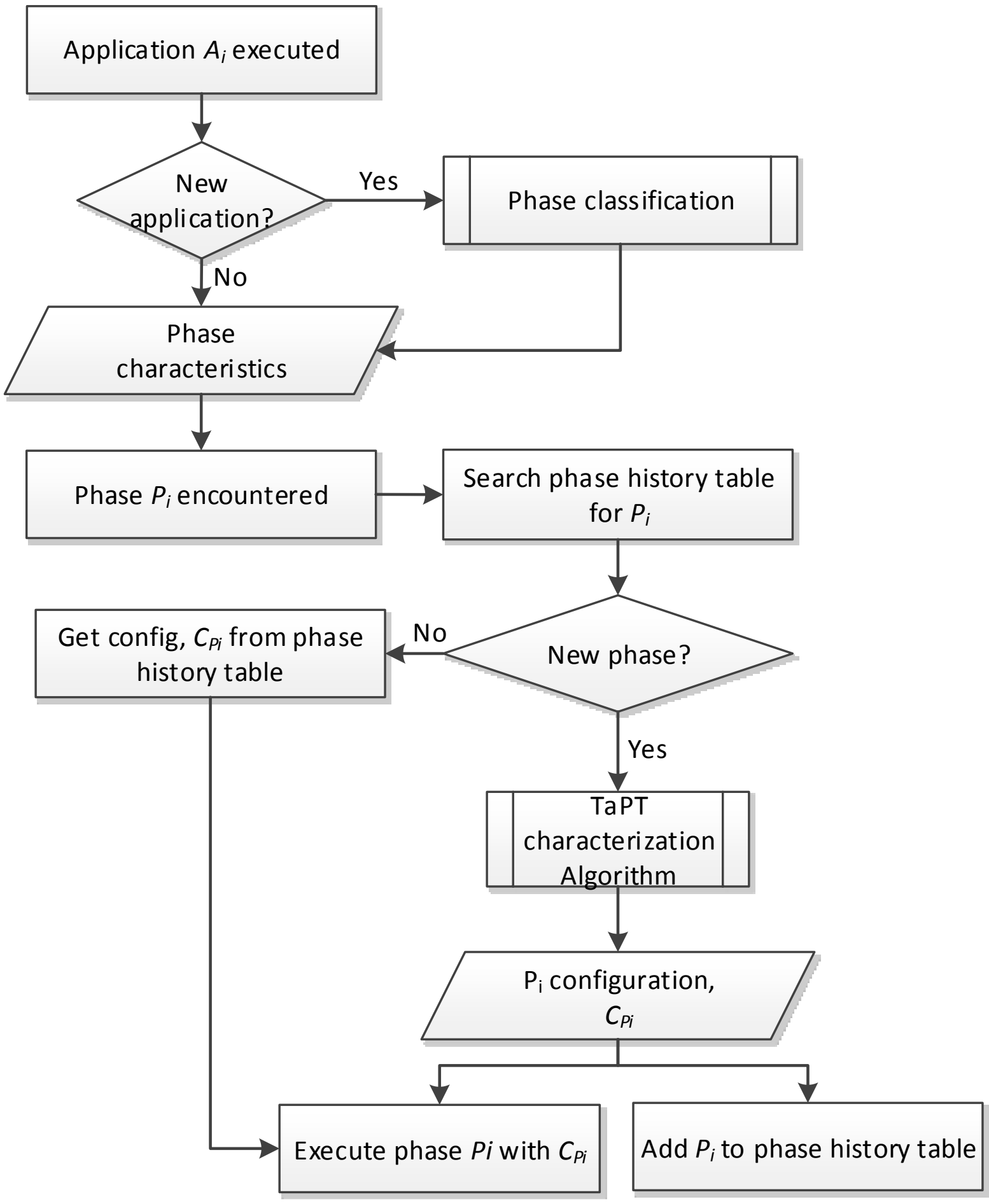

Fig. 3. TaPT overview

To minimize tuning overhead, TaPT only tunes distinct phases and uses the determined configurations for reoccurrences of that phase. A phase history table stores information about previously executed phases and the phases' best system configurations. When a phase $P_i$ is executed, if $P_i$ is in the phase history table, $P_i$ has been previously executed (i.e., $P_i$ is a not new phase) and the stored best system configuration $C_{Pi}$ is used to execute $P_i$. If $P_i$ is not in the phase history table (i.e., $P_i$ is a new phase), TaPT determines $P_i$'s best system configuration $C_{Pi}$ using the characterization algorithm (Section 3.3), $P_i$ is executed with $C_{Pi}$ and $C_{Pi}$ is stored in the phase history table for subsequent executions of $P_i$.

### 3.2 TaPT Architecture

Fig. 4 depicts the TaPT architecture for a sample dual-core system, which can be extended to any *n*-core system. The on-chip components include processing cores that are connected to the L1 caches and the TaPT module. Without loss of generalization, we assume that the L1 caches are directly connected to off-chip main memory, and since this hierarchy implies that there is no dependence between the caches, the caches can be tuned independently. We note that this is a viable assumption with respect to current state-of-the-art microprocessors [ARM 2016].

The TaPT module includes a *cache tuner,* a *DFS controller*, a *phase classification module,* and a *phase history table.* The cache tuner [Adegbija et al. 2014] and DFS controller [Herbert and Marculescu 2009] interface with the caches and processing cores to set cache configurations and clock frequencies, respectively, as determined by the TaPT algorithm (Section 3.3). The phase classification module uses execution statistics from the cores' hardware counters to classify an application's execution into phases at runtime, and the phase history table stores cache configurations and clock frequencies of previously executed phases for subsequent execution of those phases. We discuss details of the hardware overheads in Section 3.4.

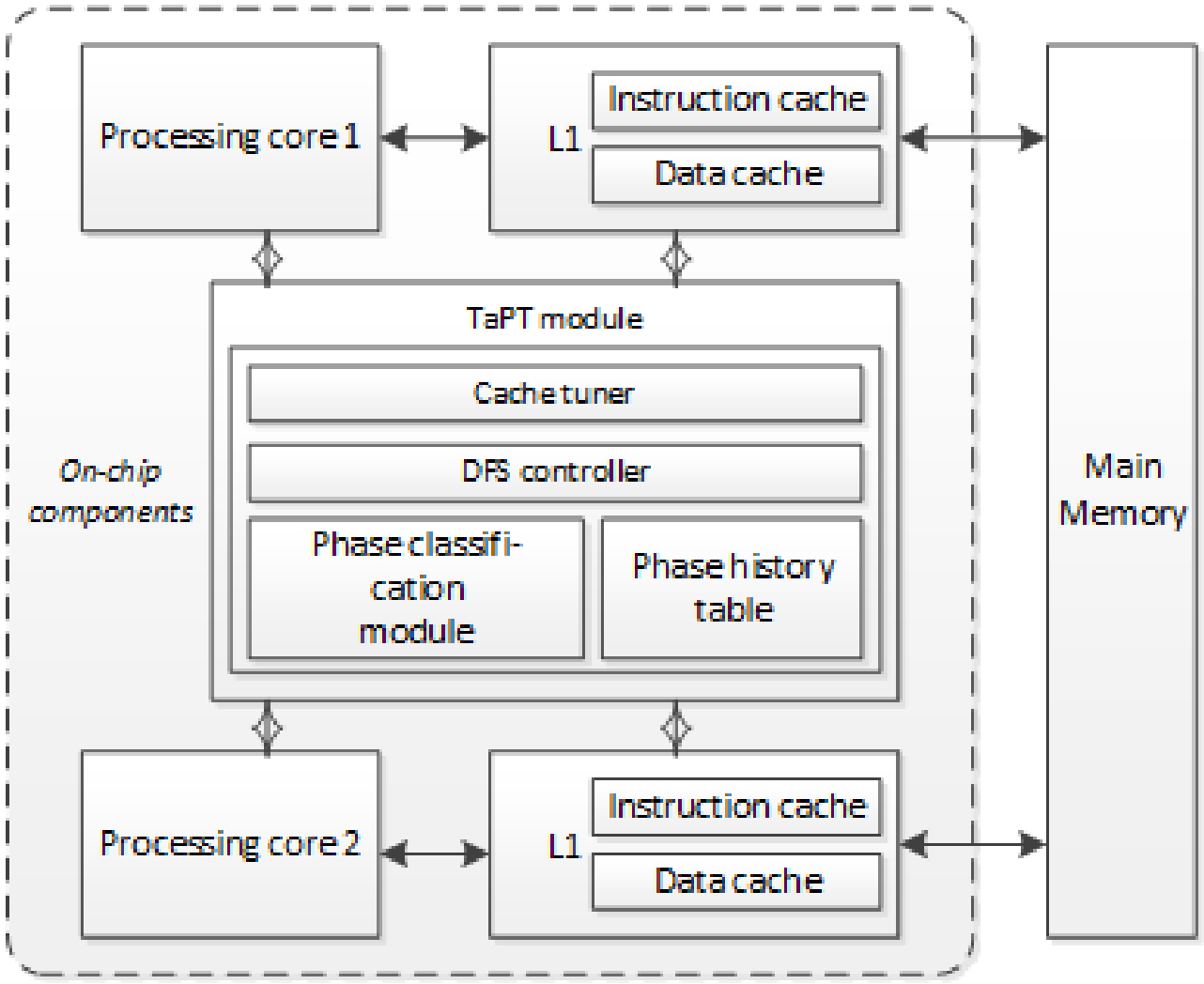

Fig. 4. TaPT architecture.

### 3.3 TaPT Characterization Algorithm

The characterization algorithm determines each phase's best configuration that emphasizes designer-specified optimization priorities. Alternatively, when no priority is specified, the algorithm determines Pareto optimal configurations that automatically emphasize the energy delay product, to account for both energy consumption and execution time, while also reducing the temperature and/or preventing significant temperature increase.

TaPT allows the designer to prioritize optimization of execution time, energy, and temperature through priority settings $X$, $N$, and $T$, respectively. When a priority setting is selected, TaPT efficiently determines the best system configuration $C_{Pi}$ for a phase $P_i$ while adhering to designer-specified constraints. The priority settings trade off the non-prioritized design objective in favor of the prioritized design objective. For example, $N$, which prioritizes energy optimization, trades off increased execution time and increased temperature for minimized energy. Alternatively, $X$, which prioritizes execution time, trades off energy consumption and temperature for reduced execution time. If the designer does not specify a priority, the priority setting defaults to $S$, which prioritizes energy delay product (EDP). To increase optimization flexibility, TaPT also allows the designer to associate a peak temperature threshold with each priority setting. Thus, when the designer specifies a temperature threshold, TaPT determines Pareto optimal configurations that do not exceed the threshold while optimizing other optimization goals.

To ensure equal probability of selection for all configurations when generating the population, TaPT uses random uniform distribution. On system startup, since there are no previously executed phases, the initial archive is an empty set. TaPT generates $P_i$'s archive from $P_i$'s population's and archive's non-dominated configurations (Equation (1)) using the configurations' fitness and stores $P_i$'s final archive in the phase history table. A configuration $C_i$'s fitness is the sum of $C_i$'s dominators' strengths, and a configuration's $C_i$'s strength $S(C_i)$ is the number of configurations dominated by that configuration such that:

$$S(C_i) = |\{C_j \mid C_j \in P \cup A \ \forall \ C_i \succ C_j\}| \quad (2)$$

where $P$ and $A$ are the $P_i$'s population and archive, respectively. $C_i$'s fitness $R(C_i)$ is:

$$R(C_i) = \sum S(C_j) \ \forall \ C_j \in P \cup A, C_j \succ C_i \quad (3)$$

where $R(C_i) = 0$ indicates that $C_i$ is non-dominated.

To implement phase-based tuning, TaPT calculates the phase distances [Adegbija and Gordon-Ross 2014] between the currently executing phase $P_i$ and all of the previously executed phases $P_{i-1}$, $P_{i-2}$, …, $P_{i-n}$. The phase distance is the difference between two phases' characteristics. In previous work, where we used phase distances for tuning cache configurations, we used the normalized difference between two phases' cache miss rates to calculate the phase distance between those phases [Adegbija and Gordon-Ross 2014]. Using a single execution characteristic (e.g., cache miss rates) to calculate phase distances suffices when only one hardware parameter (e.g., cache configurations) is being tuned. However, when multiple hardware parameters that affect multiple execution characteristics are tuned, the phase distance must be computed using a multidimentional distance metric for accurate representation.

Since TaPT tunes multiple hardware parameters (instruction and data cache configurations and clock frequency), TaPT calculates the phase distance using the Euclidean distance between the instruction cache miss rate ($iMR$), data cache miss rate ($dMR$), and the instruction per cycle (IPC). The phase distance $D$ between two phases $P_i$ and $P_j$ is:

$$D = \sqrt{(iMR_{Pi} - iMR_{Pj})^2 + (dMR_{Pi} - dMR_{Pj})^2 + (IPC_{Pi} - IPC_{Pj})^2} \quad (4)$$

TaPT uses the most similar phase's archive as the currently executing phase's initial archive. The most similar phase has the minimum $D$ from $P_i$. Since phases with stable characteristics require similar configurations, using the most similar phase's archive as $P_i$'s initial archive starts the TaPT algorithm with solutions that are presumably closer to $P_i$'s Pareto optimal solutions, as compared to an archive from the randomly-generated initial population.

Algorithm 1 depicts the TaPT algorithm, which executes for each new phase $P_i$. The algorithm takes as input the number of previously executed phases $n$ and a designer-specified population size $s$, archive size $A_{size}$, number of generations $G$, and priority setting $Q$. The algorithm outputs $P_i$'s best system configuration. The product of $s$ and $G$ defines the maximum number of configurations explored/executed during tuning, which limits the tuning overhead, and $A_{size}$ specifies the size of the archive and ensures that only the most fit configurations (Equations (2) and (3)) are stored in the archive. Given the nature of evolutionary algorithms, the archive does not necessarily contain the actual Pareto optimal solutions. In general, larger $s$ and $G$ values determine solutions that are closer to the Pareto optimal solutions, but also increase tuning overhead. Alternatively, smaller $s$ and $G$ values reduce tuning overhead, but may also determine configurations that are farther from the Pareto optimal solutions. We extensively evaluated different values of $s$, $G$, and $A_{size}$ and observed that $s$ and $G$ values that explored 4% of the design space and $A_{size}$ = 5 yielded an efficient balance between determining Pareto optimal solutions and reduced tuning overhead.

First, TaPT generates an initial population from the configuration space and calculates the phase distance $D$ between the currently executing phase and all of the previously executed phases (lines 1 – 7). Next, TaPT initializes $P_i$'s archive to $P_i$'s most similar phase's archive (i.e., the phase with the minimum distance $D$ from $P_i$) (lines 8 and 16). At system startup (n = 0), there are no previously executed phases ($D$ = null), and the archive is initialized to an empty set (lines 9 – 10). For each generation, TaPT uses the previous generation's Pareto optimal set as the current generation's initial archive (line 15). TaPT calculates each population's and archive's configuration's fitness using Equations (2) and (3), and updates the current generation's archive with the non-dominated configurations (lines 17 – 21). TaPT maintains $P_i$'s archive's size at $A_{size}$ by discarding the least fit configurations or adding the most fit configurations from the population (line 22).

**ALGORITHM 1.** TaPT Algorithm

**Input:** *n, s,* $A_{size}$, *G*, *Q*
**Output:** $P_i$'s best configuration

```
0     t ← 0
1     for i ← 1 to s do
2         C_i ← rand() / s + 1
3     end
4     population is {C_1, C_2, …, C_s}
5     for j ← 1 to n do
6         D_j ← d(P_i, P_j)
7     end
8     A_msp ← archive(P_j) | D = min(D_j)
9     if n == 0 && t == 0 then
10        archive ← ∅
11    else if k > 0 && t == 0 then
12            archive ← A_msp
13        end
14    else
15        archive ← archive(t-1)
16    end
17    U ← population + archive
18    for (C_i ∈ U) do
19        fit(C_i) ← calculateFitness(C_i)
20    end
21    archive ← getNonDominated(U)
22    size(archive) ← A_size
23    if t == (G – 1) then
24        bestConfiguration(P_i) ← min(f(Q))
25        exit
26    else
27        t++
28        goto line 1
29    end
```

On the final generation, TaPT selects the best configuration from the archive that optimizes the specified priority setting (line 24). Finally, TaPT stores $C_{P_i}$ in the phase history table (Fig. 4) for $P_i$'s subsequent executions.

### 3.4 Computational Complexity and Hardware Overhead

TaPT calculates $S(C_i)$ and $R(C_i)$ with worst-case time complexity $O(m^2)$, where $m$ is the sum of the population and archive sizes, and calculates $D$ with worst-case time complexity $O(n)$, where $n$ is the number of previously executed phases. Thus, since these calculations dominate TaPT, TaPT results in minimal computation overhead. Furthermore, TaPT utilizes previously proposed and implemented hardware, such as the phase history table [Sherwood et al. 2003] and the DFS controller [Herbert and Marculescu 2009], which have been shown to constitute little hardware overhead with respect to the microprocessor. Additionally, we have previously proposed and designed scalable and efficient cache tuners using synthesizeable VHDL and synthesized with Synopsys Design Compiler [Synopsys 2014]. Our cache tuner constitutes an average area overhead of 4.73% with respect to a MIPS32 M4K microprocessor [MIPS 2016], and we estimate that this overhead will reduce even further in larger systems. Details of our cache tuner design can be found in [Adegbija et al. 2014].

## 4. EXPERIMENTS

### 4.1 Experimental Setup

We evaluated TaPT's execution time, energy, EDP, and temperature savings by comparing a system that switches to the best configuration, as determined by TaPT, for each phase to a base system with a fixed system configuration. The base system had 32 Kbyte, 4-way private level one (L1) instruction and data caches with 64 byte line sizes, and a 2 GHz operating core frequency. This configuration is similar to current embedded systems [Motorola 2014], and thus serves as a good base comparison to a commercial off-the-shelf (COTS) system.

We modeled an embedded processor architecture, similar to the ARM Cortex A9 [ARM 2016], consisting of a 4-width out-of-order issue processor with 8 pipeline stages and 45 nm technology. Our experiments represent state-of-the-art embedded systems, and our results and analyses extend to future and/or more complex systems (e.g., $n$-core processors, heterogeneous systems, etc.) because TaPT is independent of these system characteristics. The processor's configurable L1 instruction and data caches' sizes ranged from 8 to 32 Kbyte, line sizes ranged from 16 to 64 byte, and associativities ranged from 1- to 4-way, all in power-of-two increments. The processor offered seven clock frequencies ranging from 800 MHz to 2 GHz in 200 MHz increments. Given these parameter values, the design space contains 1,701 configurations.

We modeled the processor using GEM5 [Binkert et al. 2011] and generated cache miss rates and core statistics, which we used to calculate the execution time. We also used these statistics to calculate the system's total energy consumption and EDP with McPAT [Li et al. 2009]. We used Hotspot 5.0 [Skadron et al. 2004] as the thermal modeling tool to measure the temperature using a floorplan and silicon chip area similar to the ARM Cortex A9 processor. We ran thermal simulations and sampled the application's power consumption at 10 ms intervals, similar to modern operating systems (e.g., Linux) [Sharafi et al. 2010]. Previous work showed that this fine-grained sampling accurately depicted the application's temperature characteristics during execution [Sharafi et al. 2010]. To simulate an embedded system without cooling mechanisms, such as a heat sink and/or spreader, we set the convection resistance to 4K/W and the heat sink and spreader thickness to 1 mm and 0.1 mm, respectively, which are considered negligible in Hotspot.

To model a variety of real-world embedded system applications, we used seventeen benchmarks: twelve EEMBC [Poovey et al. 2009] Automotive benchmarks (the full suite could not be evaluated due to compilation errors) and five MiBench [Guthausch et al. 2001] benchmarks selected to represent different application domains. The benchmarks were specific compute kernels performing specific tasks in different application domains, such as networking, image processing, security, etc.

We implemented TaPT using Perl scripts to drive simulations and executed each phase once to completion. To implement phase classification, we ran execution trace simulations on each benchmark using GEM5 to generate cache miss rates and IPC statistics, and grouped intervals with similar characteristics as phases using variable-length intervals, which previous work found to be effective for phase classification [Gordon-Ross and Vahid 2003]. Since the benchmarks were specific compute kernels, our experiments revealed that the benchmarks exhibited relatively stable characteristics throughout execution. Without loss of generality, this characteristic stability enabled us to consider each kernel/benchmark as a different phase of execution.

### 4.2 Temperature Impact of Cache Configurations

To evaluate the impact of variable cache configurations on the system's temperature, we extensively analyzed the temperature variations over exhaustive executions of the cache configurations in our design space (Section 4.1). We executed all seventeen benchmarks for 240 possible cache configurations. For brevity, we only show results for exhaustive executions from one benchmark each from the EEMBC and MiBench benchmark suites – *a2time01* and *sha,* respectively. The benchmarks represent variations of compute intensity, where *a2time01* and *sha* represent applications with low and high compute intensity, respectively, however, we observed similar trends across the other benchmarks.

Fig. 5 (a) and (b) depict the temperature variations over several cache configurations for *a2time01* and *sha* with the clock frequency set at 2 GHz. In general, the peak temperatures ranged from 69$^{o}$C to 90$^{o}$C, with a standard deviation of 5. As expected, the base configuration, with the largest cache configuration had the highest temperature. Since the impact order of cache parameters on energy consumption has been widely used for developing cache tuning heuristics for energy consumption [Zhang et al. 2003], we also analyzed the trend in temperature changes as the cache configurations changed. We observed that the temperature changes did not follow the same trend as energy changes. This observation implies that previous knowledge on the impact order of cache parameters on energy consumption does not apply to temperature optimization.

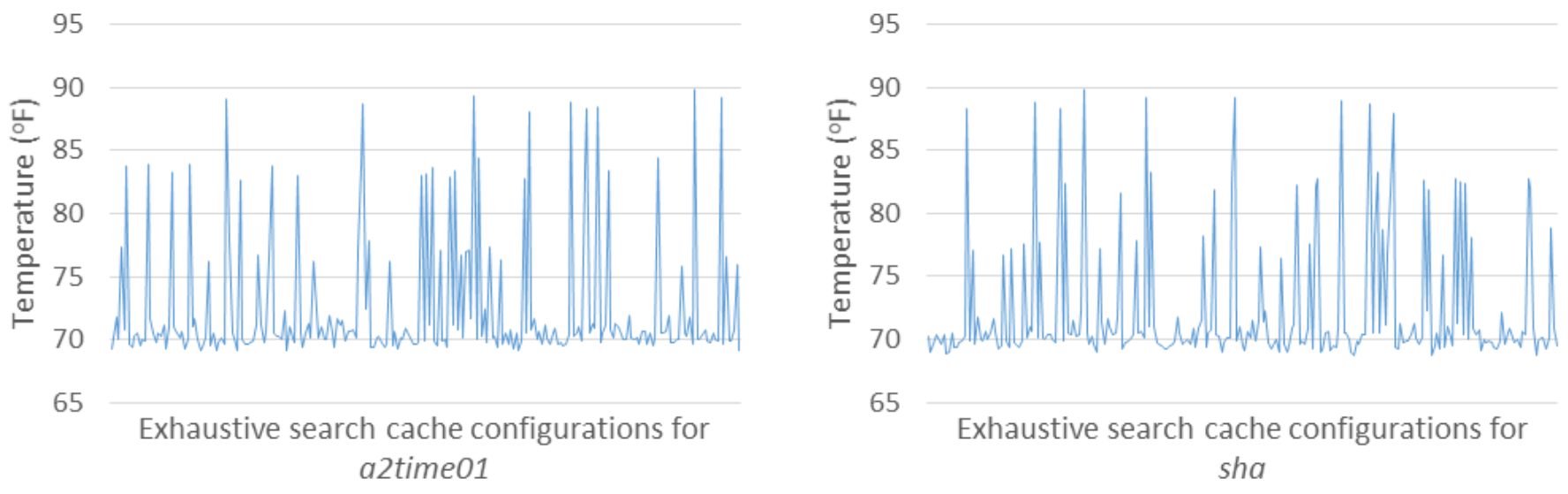

*(a)* *(b)*

Fig. 5. Temperature variations for different cache configurations

Thus, we explored how changes in cache parameter values (cache size, line size, and associativity) affected the temperature with respect to the base configuration. To provide a clearer picture of the impact of each cache parameter value, we varied each parameter value while keeping the others constant. Fig. 6 shows the temperatures when the cache parameter values were changed from the base configuration. Surprisingly, reducing the cache size from 32KB to 16KB and 8KB only reduced the temperature by 1.1°C (1% reduction) and 2°C (1.7% reduction), respectively. While these apparently insignificant reductions may be impactful over a long period of time, the small change was unexpected, especially since the cache size usually has the largest impact on other optimization goals, such as energy and execution time.

Unlike the cache size, the associativity and line size had much larger impacts on the temperature. Reducing the associativity from 4-way to 2-way and 1-way (with all other parameters at the base values) reduced the temperature by 15.7°C (18% reduction) and 15.8°C (18% reduction), respectively. Reducing the line size from 64B to 32B and 16B reduced the temperature by 16.9°C (19% reduction) and 17.5°C (20% reduction), respectively. Thus, our results reveal that the cache line size has the largest impact on temperature, followed by the associativity, and then the cache size. The parameter with the largest impact would likely be the best to tune first. Furthermore, this insight can inform the tradeoffs involved in developing heuristics that tune multiple parameters. We intend to explore these insights further in future work.

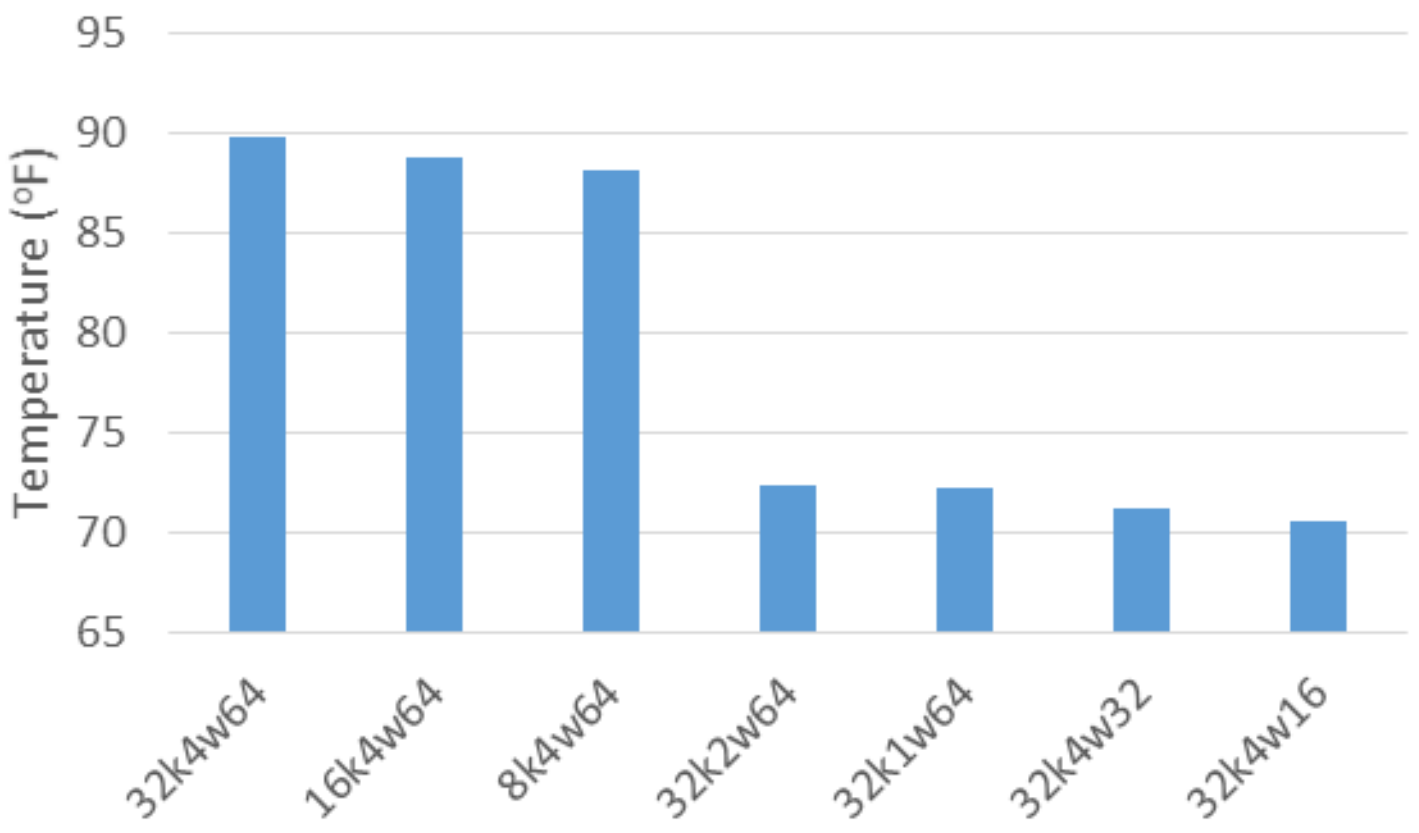

Fig. 6. Impact of cache parameter values on temperature

### 4.3 TaPT Parameters

To determine appropriate values for $s$, $G$, and $A_{size}$, we conducted a sensitivity study to quantify the impacts of different values of $s$, $G$, and $A_{size}$ on optimization efficiency. We exhaustively explored a design space comprised of $s$ ranging from 10 to 100 in increments of 10, and $G$ and $A_{size}$ ranging from 3 to 6 in increments of 1, resulting in a total of 160 possible

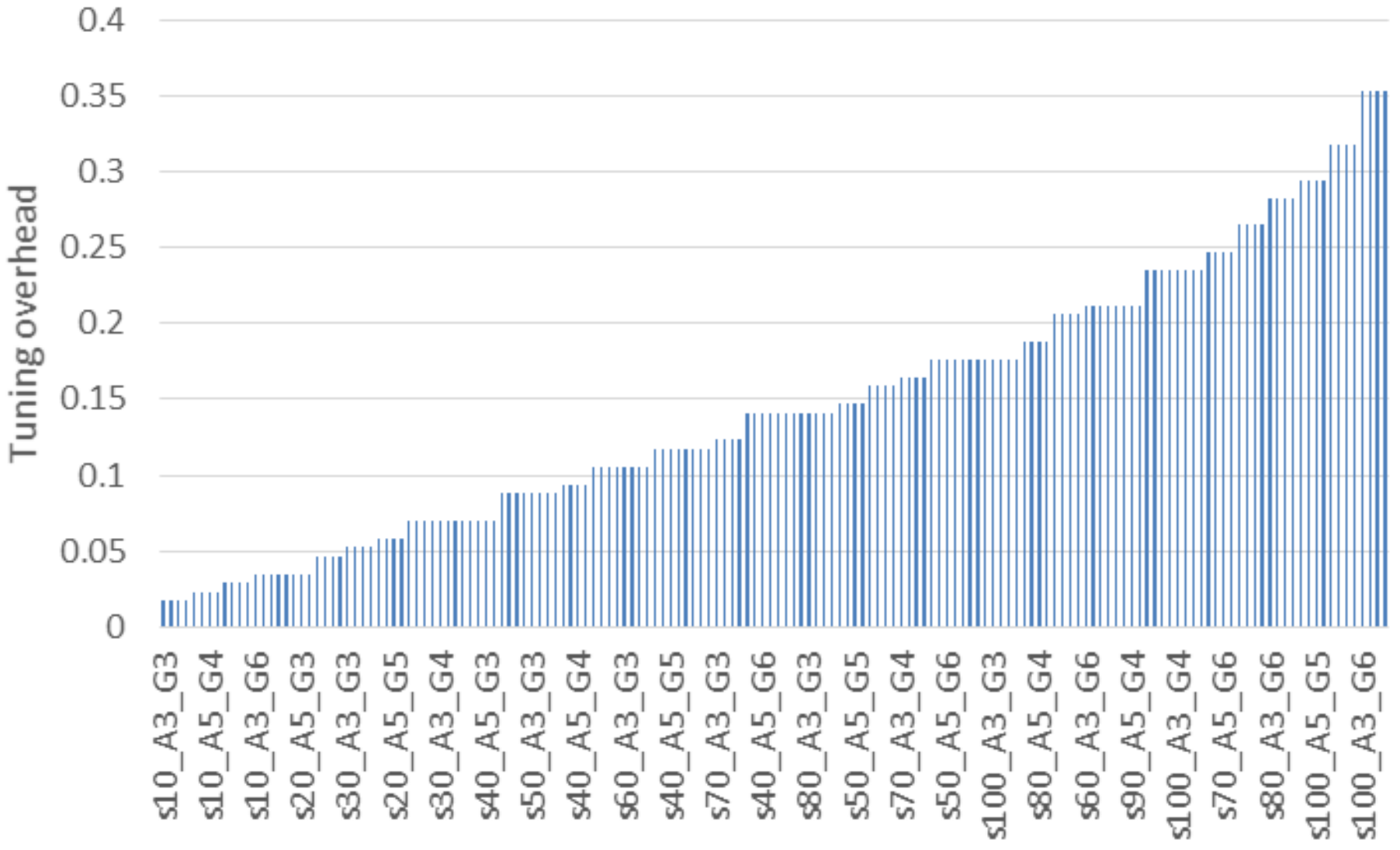

*(a)*

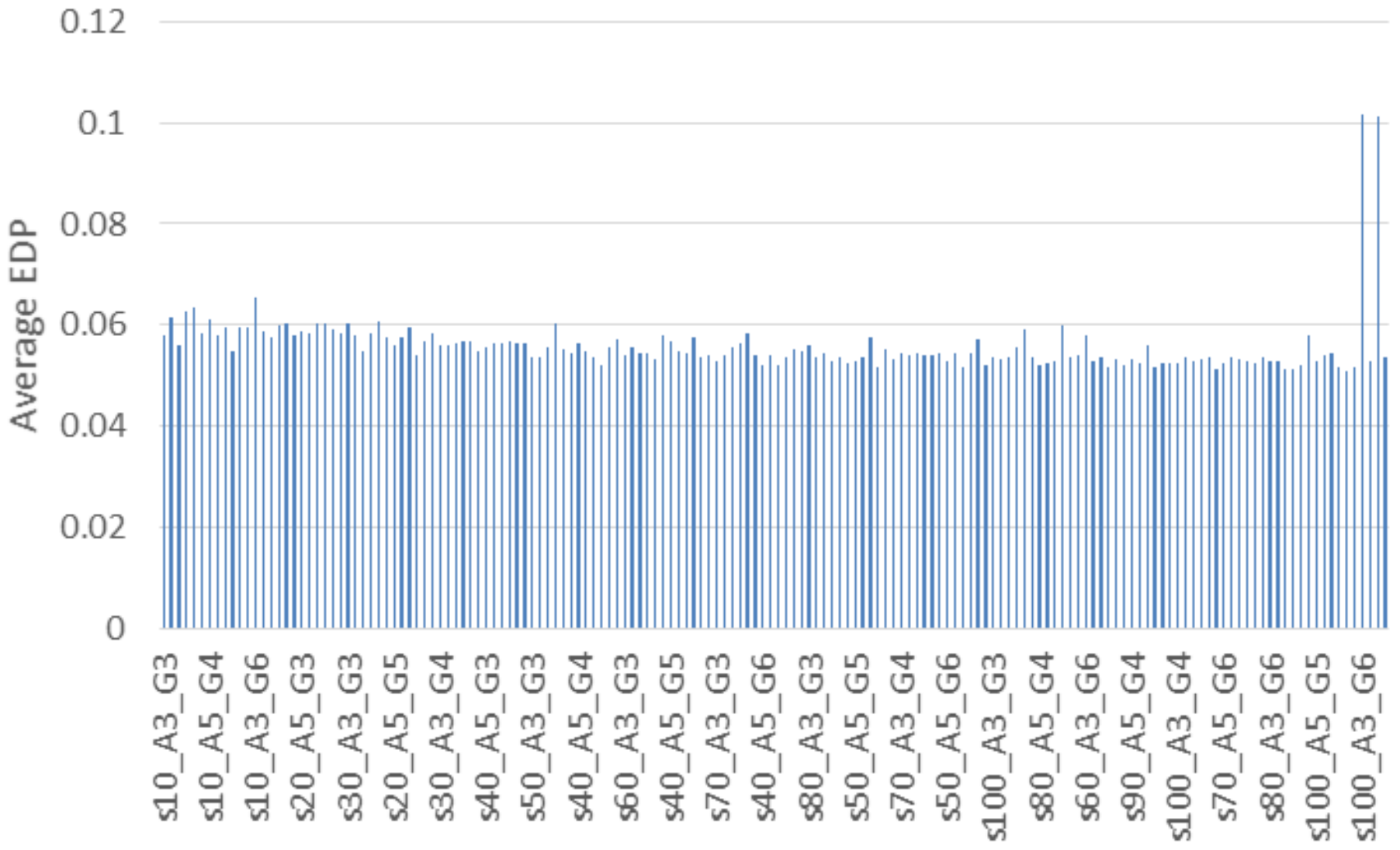

*(b)*

Fig. 7 (a) Tuning overhead incurred and (b) Average EDP achieved by different TaPT parameters combinations. To represent optimization efficiency, we used the average energy delay product (EDP) achieved by the TaPT parameters while executing all our experimental benchmarks.

Fig. 7 (a) and (b) illustrate the relationships between the TaPT parameters, and the tuning overhead and EDP, respectively. The x axis (on both figures) represents the TaPT parameters, denoted as 's*x*_A*y*_G*z*', where *x, y,* and *z* represent the population size, number of generations, and archive size, respectively. For brevity, not all the TaPT parameters are shown on the figures, however, the shown parameters are representative of all the TaPT parameters. Fig. 7 (a) shows that the tuning overhead increased steadily for the TaPT parameters, ranging from 2% to 35% tuning overhead. However, Fig. 7 (b) shows that an increase in tuning overhead did not necessarily result in EDP reduction. The EDP achieved by different TaPT parameters did not change significantly as the design space increased. Thus, we performed additional analysis to identify TaPT parameters that achieved sufficient tradeoffs between tuning overhead and optimization efficiency.

Our analysis revealed that s = 20, G = 3, and $A_{size}$ = 5 achieved a good balance between Pareto optimal solutions and tuning overhead. These values explored less than 4% of the design space, while larger values increased tuning overhead without significantly improving the Pareto optimal solutions and smaller values reduced tuning overhead, but achieved sub-Pareto-optimal solutions. *s* and *G* are system dependent and can be scaled appropriately for different design spaces.

To explore several diverse design objectives, we modeled all of TaPT's priority settings using these values of *s*, *G*, and $A_{size}$. To evaluate the impact of designer-specified temperature thresholds lower than the base configuration's average peak temperature of 89°C (determined by simulation), we evaluated empirically-determined high and low temperature thresholds set at 82°C and 65°C, based on the range of temperatures observed during simulation. The high 82°C threshold illustrates a system where the primary concern is for the temperature to be maintained below 82°C to prevent overheating damage, while the low 65°C threshold represents a strict temperature-constrained system to illustrate how maintaining a low temperature impacts the other objective functions.

### 4.4 TaPT Optimization Results

Fig. 8 and Fig. 9 depict the execution time, energy, EDP and temperature of the best configurations as determined by TaPT normalized to the base system configuration for a single execution of each benchmark/phase for each priority setting [Adegbija and Gordon-Ross 2014]. Fig. 8 (a) illustrates the optimization benefits from TaPT's default setting of *S* (EDP prioritization), which involves no designer effort and no specified temperature threshold. On average over all the applications, the EDP, energy, execution time, and temperature reduced by 31%, 30%, 2%, and 21%, respectively, with maximum reductions of 48%, 35%, 19%, and 5%, respectively. For some phases, prioritizing EDP minimization only slightly reduced the temperature. For example, *candr01*'s EDP, energy, and execution time reduced by 40%, 27% and 18%, respectively, while reducing the temperature by only 8%. However, prioritizing EDP minimization increased the execution time for other phases by up to 6%, but gained significant reductions in energy and temperature. For example, *mad*'s EDP, energy, and temperature reduced by 23%, 26%, and 21%, respectively, while increasing the execution time by 4%. In general, priority setting $S$ minimized the EDP, and reduced the energy consumption and temperature for all phases, with only minor increases in execution time for some phases.

Fig. 8 (b) depicts the average execution time, energy, EDP, and temperature savings for a priority setting $N$ (energy prioritization) and an 82°C temperature threshold. The execution time, energy, EDP, and temperature reduced by 4%, 31%, 34%, and 20%, respectively. We also evaluated TaPT's optimization behavior when a temperature threshold is specified. We evaluated this behavior using an 82°C temperature threshold. Fig. 10 depicts the phases' peak temperatures with respect to the threshold temperature of 82°C. TaPT maintained the temperature at or below 82°C for all the phases, but did not minimize the temperature. The 82°C threhold allowed for greater execution time, energy, and EDP reduction a system that prioritized the temperature or one with a lower temperature threshold. This option is suitable for situations where the temperature constraints are known and can be specified by the designer.

To illustrate TaPT's optimization capabilities with a low temperature threshold and priority setting $T$ (prioritize temperature), Fig. 9 (a) depicts the execution time, energy, EDP, and temperature savings with a 65°C temperature threshold and priority setting $T$. On average, over all the phases, the energy and temperature decreased by 13% and 25%, respectively. However, the execution time and EDP significantly increased by 39% and 22%, respectively. TaPT maintained a peak temperature for all

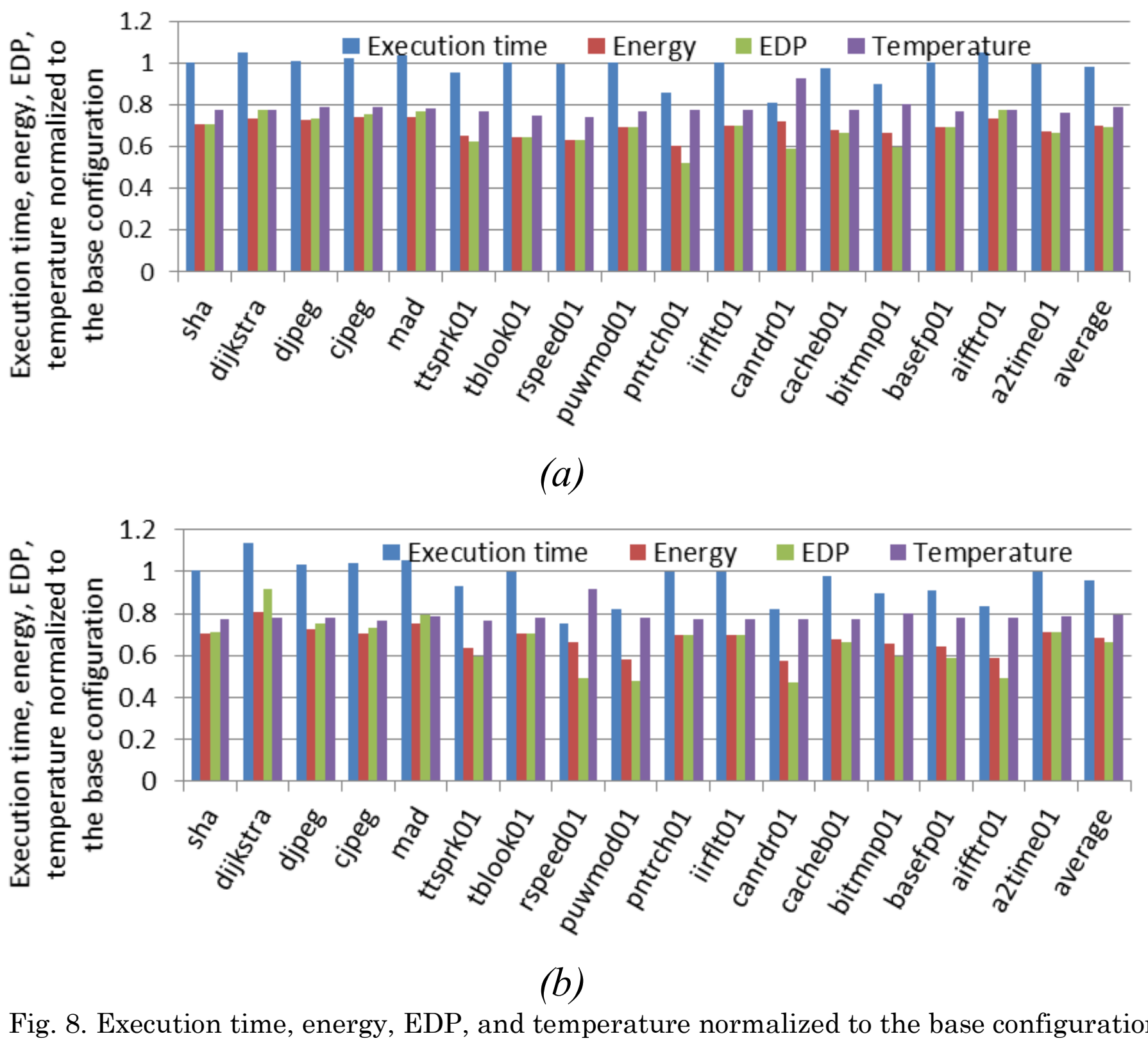

Fig. 8. Execution time, energy, EDP, and temperature normalized to the base configuration for priority settings (a) $S$ and (b) $N$.

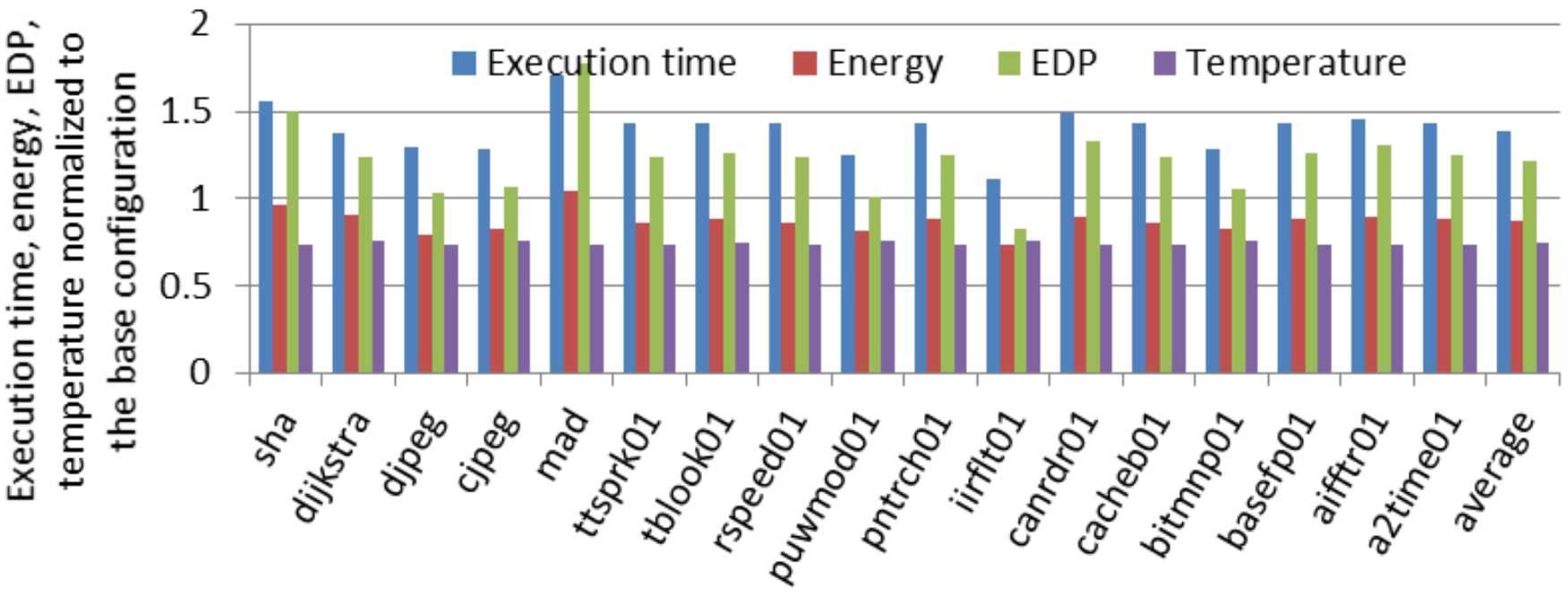

*(a)*

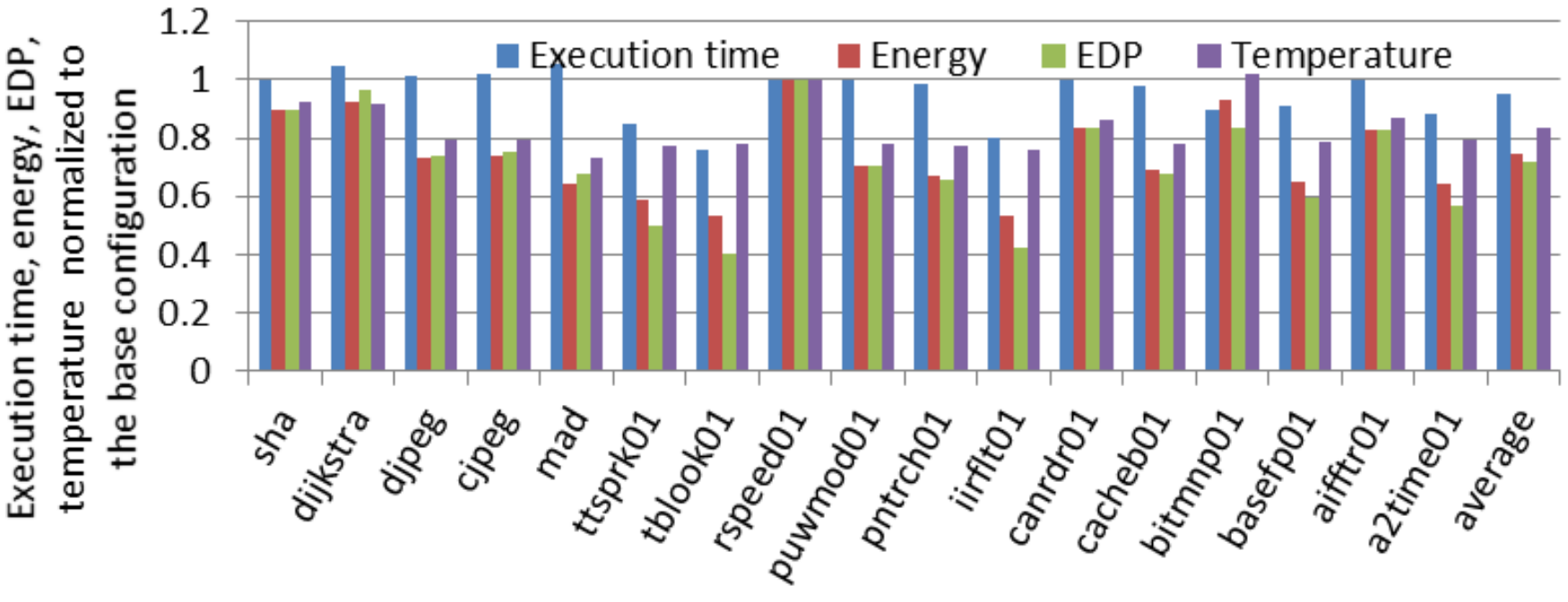

*(b)*

Fig. 9. Execution time, energy, EDP, and temperature normalized to the base configuration for priority settings (a) *T* and (b) *X*.

the phases within 65 to 68°C, however, to maintain this low peak temperature, TaPT traded off execution time and energy consumption. Increasing the temperature threshold to 70°C (results not shown for brevity) decreased the energy, EDP, and temperature by 27%, 26%, and 21%, respectively, while the execution time only increased by 2%. These results show TaPT's ability to trade off optimization goals in order to adhere to design constraints. The results also show the extent to which some optimization goals may be adversely affected in a multi-objective optimization problem where one of the objective functions is significantly constrained.

Fig. 9 (b) shows that when using priority setting *X* (execution time prioritization) with no temperature threshold, TaPT reduced the execution time, energy, EDP, and temperature by 5%, 26%, 29%, and 16%, respectively. For some phases, the reductions were more significant. For example, TaPT significantly reduced *tblook01*'s execution time, energy, EDP and temperature by 24%, 47%, 60%, and 22%, respectively. However, for some phases the execution time did not reduce. For example, *mad*'s execution time increased by 5% while the energy, EDP, and temperature decreased by 36%, 32%, and 27%, respectively. We observed that even though TaPT achieved significant execution time improvement for some phases, the base configuration was the best for execution time optimization for most phases. Thus, TaPT attempted to reduce energy and temperature without significantly degrading the execution time. Overall, TaPT succeeded in trading off optimization goals, where necessary, in order to satisfy designer specified optimization priorities, without significantly degrading the other optimization goals.

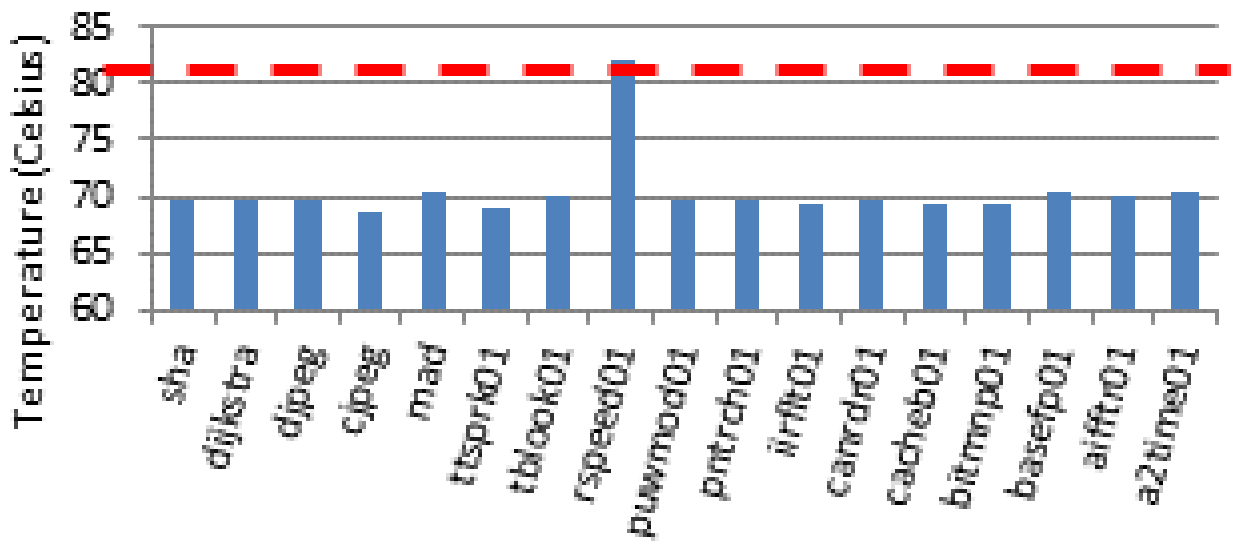

Fig. 10. Peak temperatures with respect to a temperature threshold of 82°C (broken horizontal lines).

### 4.5 Comparison to Prior Work

To further evaluate TaPT's effectiveness, we compared TaPT to prior work using DFS or cache tuning in isolation. For both DFS and cache tuning, we used exhaustive search to determine the best configurations for each benchmark to represent the optimal configurations (i.e., best-case optimization scenarios). Similar to the previous experiments, we assumed priority settings, *S, N, T,* and *X,* where the prioritized setting was the optimal for both DFS and cache tuning.

Fig. 11 depicts the average execution time, energy, EDP, and temperature of the best configurations as determined by TaPT normalized to the best DFS configurations (best frequency), using the base cache configuration for priority settings *S, N, T,* and *X*. With the default priority setting *S,* TaPT reduced the average EDP, energy, execution time, and temperature by 41%, 39%, 38%, and 46%, respectively, as compared to DFS. With priority setting *N,* TaPT reduced the average EDP, energy, execution time, and temperature by 34%, 32%, 29%, and 24%, respectively, as compared to DFS. With priority setting *T,* TaPT reduced the average

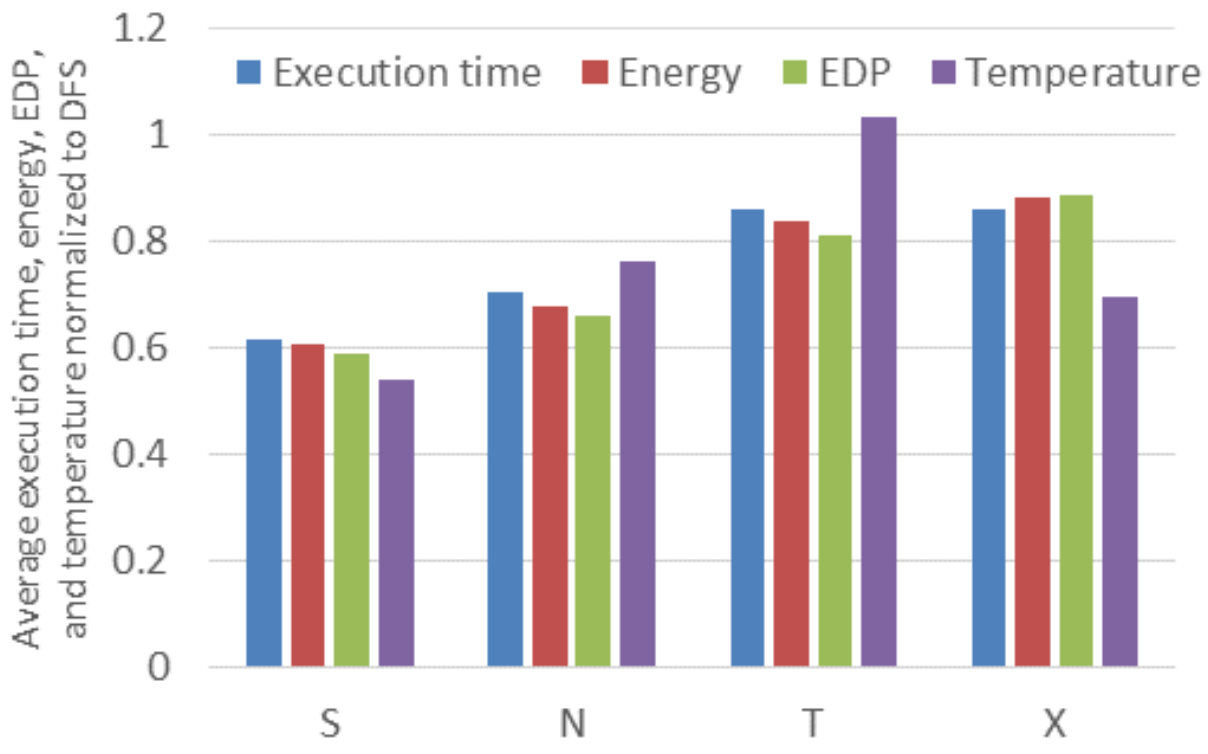

Fig. 11. Average execution time, energy, EDP, and temperature normalized to DFS for priority settings *S, N, T,* and *X.*

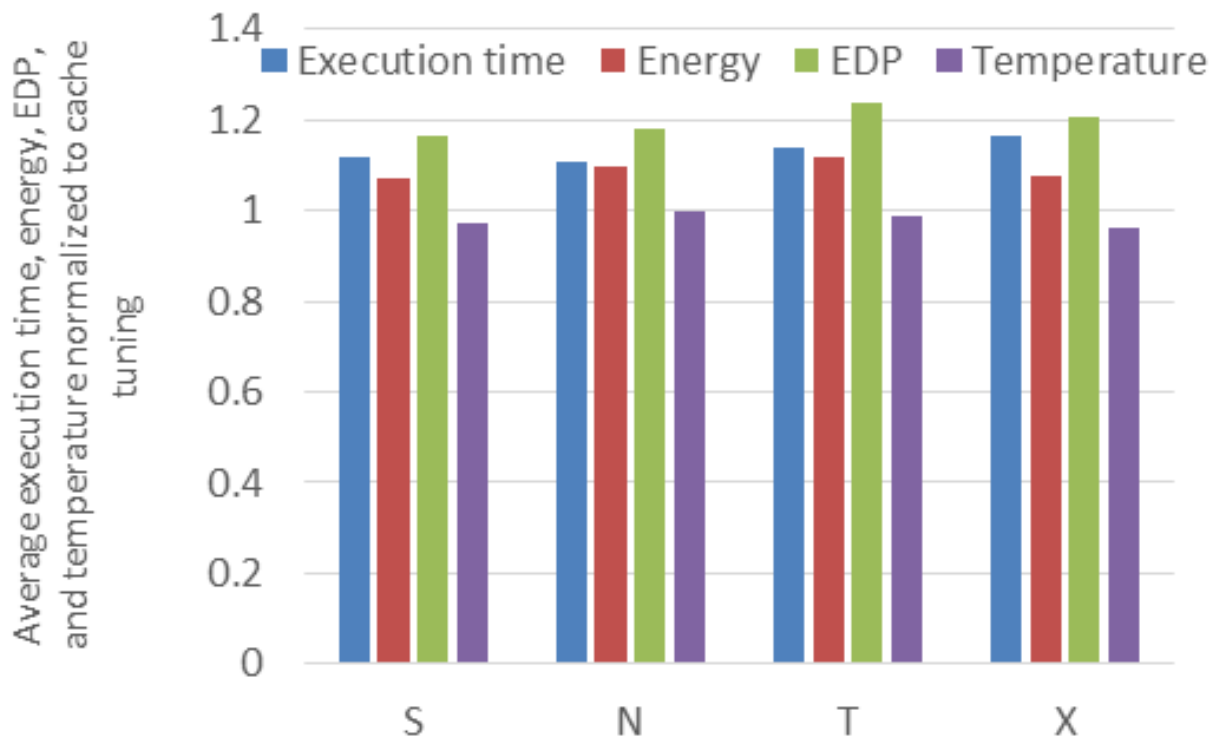

Fig. 12. Average execution time, energy, EDP, and temperature normalized to cache tuning for priority settings *S, N, T,* and *X*.

EDP, energy, and execution time by 19%, 16%, and 14%, respectively, as compared to DFS. However, TaPT *increased* the average temperature by 3% as compared to DFS. This temperature increase was due to TaPT's optimization of other design objectives while prioritizing the temperature. DFS achieved lower temperature than TaPT because of the significant impact of the clock frequency on temperature. Finally, with priority setting *X,* TaPT reduced the average EDP, energy, execution time, and temperature by 11%, 11%, 14%, and 31%, respectively, as compared to DFS. These results illustrate TaPT"s ability to optimize multiple objectives while prioritizing the designer's selected priority setting.

Fig. 12 depicts the average execution time, energy, EDP, and temperature of the best configurations as determined by TaPT normalized to the best cache configurations as determined by exhaustive search, using the base clock frequency for priority settings *S, N, T,* and *X*. Unlike when comparing to DFS, cache tuning outperformed TaPT in all priority settings for all optimizations except temperature optimization. We note that this result was expected since we used exhaustive search to determine the best configurations from the complete design space afforded by cache tuning. These configurations represent the best-case scenarios and do not reflect a real-world multi-objective optimization scenario. With priority setting *S,* TaPT increased the average EDP, energy, and execution time by 16%, 7%, and 12%, respectively, and reduced the temperature by 3%, as compared to cache tuning. With priority setting *N*, TaPT increased the average EDP, energy, and execution time by 18%, 10%, and 10%, respectively, and did not change the temperature, as compared to cache tuning. With priority setting *T*, TaPT increased the average EDP, energy, and execution time by 24%, 12%, and 14%, respectively, and reduced the temperature by 1%, as compared to cache tuning. Finally, with priority setting *X,* TaPT increased the average EDP, energy, and execution time by 20%, 8%, and 16%, respectively, and reduced the temperature by 4%, as compared to cache tuning. Thus, TaPT determined relatively similar configurations as the optimal cache tuning configurations while exploring only 4% of the design space.

### 4.6 Tuning Overhead

TaPT's tuning overhead comprises of the time it takes to determine the best cache configuration and clock frequency, and the time it takes to switch to the determined configurations. Specifically, we computed the tuning overhead in terms of the total cache tuning time, the cache configuration overhead, frequency tuning time, and the DFS transition delay overhead, which is the time it takes to switch from one frequency level to another. We assumed an average transition delay overhead of 18.24µs, similar to that of the ARM Cortex A9 [Park et al. 2013]. On average over all the benchmarks, TaPT's tuning overhead was 0.145 seconds. Due to the brief duration of most of the benchmarks, and our 10 ms tuning interval, TaPT required multiple iterations to determine the Pareto optimal configurations for most of the benchmarks. However, in embedded systems with persistent applications that execute multiple times throughout the systems' lifetime, this tuning overhead amortizes very rapidly.

To put TaPT's tuning overhead in perspective, we also compared TaPT's tuning overhead with how much time was required to determine the benchmarks' Pareto optimal configurations using exhaustive search of the design space. On average over all the benchmarks, TaPT reduced the tuning overhead from 3.62 seconds to 0.145 seconds, representing an average reduction of 96% or a 25X tuning speedup.

## 5. CONCLUSIONS

Phase-based tuning specializes a system's tunable parameters to the varying runtime requirements of an application's differerent execution phases in order to meet optimization goals, which typically involve minimizing energy consumption and/or maximizing performance. However, due to embedded systems' resource constraints, and the absence of dedicated cooling mechanisms, temperature is a growing issue in these systems. Several dynamic thermal management techniques, such as dynamic frequency scaling (DFS), task migration, etc., have been used for managing embedded systems' temperature. However, these techniques could adversely affect other optimization objectives, such as energy consumption and/or performance.

In this paper, we extensively analyzed the impacts of different cache configurations on system temperature, and showed that the cache parameters' impacts on temperature differs from the impacts on other optimization goals (e.g., energy and execution time). Our analysis revealed that the line size has the largest impact on temperature, followed by the associativity, and then the cache size. We also presented temperature-aware phase-based tuning, TaPT, which combines phase-based cache tuning and dynamic frequency scaling (DFS) to determine Pareto optimal configurations for different application phases of execution. We show TaPT's effectiveness in determining Pareto optimal configurations that significantly reduce execution time, energy, EDP, and temperature, with minimal computational complexity and low hardware overhead, while adhering to specified design constraints. Results reveal that TaPT reduces execution time, energy consumption, and temperature by up to 5%, 30%, and 25%, respectively. We also show that TaPT is easy to implement, constitutes minimal hardware overhead, and can be seamlessly incorporated in resource-constrained embedded systems.

For future work, we plan to verify TaPT's scalability to more complex systems with much larger design spaces (e.g., heterogeneous multi-/many core systems). We also intend to develop additional low-overhead heuristics that leverage the insights developed in this work for multi-objective optimization.

## ACKNOWLEDGMENTS

This work was supported in part by the National Science Foundation (CNS-0953447). Any opinions, findings, and conclusions or recommendations expressed in this material are those of the authors and do not necessarily reflect the views of the National Science Foundation.